\documentclass[12pt,preprint]{aastex}
\begin{document}
\title{The Blue Straggler Population of the Globular Cluster M5: Comparison with M3}
\author{Steven R. Warren, Eric L. Sandquist}
\affil{Department of Astronomy, San Diego State University, San Diego, CA 92128;warren@sciences.sdsu.edu;erics@sciences.sdsu.edu}
\and
\author{Michael Bolte}
\affil{UCO/Lick Observatory, University of California, Santa Cruz, CA 95064;bolte@ucolick.org }

\begin{abstract}
  
  We have surveyed the blue straggler star population of the Galactic globular
  cluster M5 using high-resolution images of the core along with wide-field
  ground-based images reaching to more than 19 core radii. To gauge M5's
  relative efficiency of producing stragglers, we compared our sample to five
  studies of other globular clusters (mainly \citealt{fer97}; \citealt{fer03};
  and \citealt{pio04}). Using a ``bright'' sample selected in the same way as
  \citet{fer97}, we found a bimodal radial distribution similar to those found
  in three other luminous clusters. When the radial distributions for
  different clusters are scaled using the core radius, there is good
  cluster-to-cluster agreement in the size of the core straggler sample and
  the center of the ``zone of avoidance''. However, M5 has the smallest
  fraction of stragglers in the zone of avoidance of any of the clusters
  measured to date, and its zone of avoidance appears to be wider (in $r /
  r_c$) than that of M3, which has a very similar surface brightness profile.
  Both of these facts indicate that M5's straggler population has dynamically
  evolved to a larger extent than M3.  Using an ultraviolet sample from {\it
    Hubble Space Telescope} selected in the same way as \citet{fer03} and
  \citet{fer04}, we find that the frequency of blue stragglers in M5 is lower
  than all but two of the clusters examined.  We also identified seven bright
  blue stragglers that were previously misidentified as HB stars by
  \citet{sand04}.  These bright stragglers are most likely the result of
  stellar collisions involving binary stars.
\end{abstract}


\section{Introduction}

Blue straggler stars (BSSs) were first identified as stars brighter and bluer
than the main-sequence turnoff in the cluster M3 by \citet{sand53}.  Since
then they have been found in all globular clusters that have been surveyed
adequately and in many open clusters. The current leading explanations
for the production of BSSs involve mass transfer within a binary system or the
merger of two stars in a binary, and stellar collisions. 

The rate of stellar collisions within a cluster depends on its
current structural characteristics (notably, central density and velocity
dispersion), and on the collision cross sections of the
interacting objects. The binary star population in a cluster can play a large
role in BSS formation.
In addition to evolution of a binary via mass-transfer and merger into a BSS, 
binaries have an enhanced collision rate for 
interaction with other stars.  BSSs, being much like newly-created main
sequence stars, are also likely to have lifetimes that are comparable to or
larger than the timescales on which a cluster dynamically evolves.  The
stragglers are thus an easily accessible population of stars that can give us
insight into a cluster's dynamical past and the importance of its present
structural characteristics.

A large amount of effort has recently been put into seeking a global view of
how BSS populations vary from cluster to cluster. \citet{pio04}
cataloged almost 3000 BSSs in the cores of 56 globular clusters based on
Hubble Space Telescope  (HST) observations,
and found a significant anticorrelation between the relative frequency of
BSS and total cluster luminosity, and little or no correlation with other
cluster parameters. Although the trend with luminosity is strong, there is a
significant amount of scatter (perhaps by as much as an order of magnitude)
around the trend.

In M3, \citet{fer93} discovered a bimodal distribution of
BSSs with respect to radius, with separate groups of core and envelope stragglers
(see also \citealt{fer97}). Since then, such distributions have been
identified in other clusters (M55, \citealt{zag}; 47 Tuc, \citealt{fer04}; NGC
6752, \citealt{sabbi}). However, in the low-luminosity cluster Palomar 13
\citep{clark}, the frequency of BSSs does not increase in the
outskirts of the cluster.

The small number of studies of the complete radial distribution of BSSs has
led us to examine the cluster M5. M5 has structural characteristics that are
currently very similar to M3 (see Table \ref{m3m5}). As such, M5 provides an
interesting test of whether clusters with similar structural properties must
necessarily have similar straggler populations.

In \S 2, we present the datasets that were used in the identification of
stragglers in M5. In \S 3, we present the total sample of stragglers, and in
\S 4, we compare the M5 sample to those from other globular clusters. In \S 5,
we discuss the comparison population of horizontal branch (HB) stars. Finally,
we discuss the size of the straggler population and its radial distribution in
\S 6.

\section{Observations and Data Reduction}

The photometric data we used in this study were derived from images taken
using several ground-based telescopes and 
HST.  Wide-field images in $B$, $V$, and $I$ were taken at the Cerro Tololo
Interamerican Observatory (CTIO) 4 m telescope, and their reduction has
already been described in \citet{sand96}. The images were centered on the
cluster core and covered square fields 16\farcm3 on a side. We used additional
higher resolution images of the cluster core from the Canada-France-Hawaii 3.5
m Telescope (CFHT) in $B$ and $I$. These images have also been discussed
previously \citep{sand96,sand04}. The field covered by the CFHT images is
approximately 2\farcm2 on a side and is slightly offset from the cluster
center, although the center is in the field.

We also made use of several HST datasets. \textit{BV} data were 
taken from the
public online data archive of \citet{pio02}. Additional unreduced images in
the ultraviolet were taken from the HST archives (proposal ID: 6607, P.I.
Ferraro.). The UV images were reduced using version 1.1.5b of the HSTPHOT
package \citep{dolphin}, which is optimized for the reduction of WFPC2 images.
The images were preprocessed using the {\it mask, getsky, crmask,} and {\it
  hotpixel} routines to eliminate saturated, vignetted, and hot pixels as well
as those with cosmic ray hits. If the images in a set were dithered, initial
guesses for the pixel offsets relative to a reference frame were taken from
WFPC2 Associations
webpage\footnote{http://archive.eso.org/archive/hst/wfpc2\_asn} in order to
align the images before cosmic ray rejection. The {\it hstphot} routine was
then used to conduct PSF photometry using precomputed
PSFs\footnote{http://purcell.as.arizona.edu/hstphot/}. We used flight-system
magnitudes in subsequent analysis of this dataset.  Positions derived from HST
images were transformed to right ascension and declination using the IRAF task
{\it metric}. These coordinates were used as the basis for the positions
[relative to the cluster center given by \citealt{har96}: $\alpha(2000) =
15^{\mbox{h}}18^{\mbox{m}}33\fs8, \delta(2000)=+2\degr 4\arcmin 58\arcsec$]
presented in later tables.

\section{The BSS sample}

We attempted to obtain a reliable global BSS sample by merging datasets for
the core of the cluster (two HST datasets and the CFHT dataset) with $r \la
150\arcsec$, and the CTIO dataset for the outskirts of the cluster $(r >
150\arcsec)$. See Table \ref{CBSS} for an example of the photometry for the
candidate BSSs.

Because there is not a reliable and uniform set of photometry covering all of
the stars in the cluster, by necessity we had to pay careful attention to our
selection criteria so that the samples in different parts of the cluster are
drawn from consistent portions of the CMD. In this section we focus on the
selection of a reliable set of bright blue stragglers separately for the 
cluster core and outskirts.

\subsection{The Core Sample \label{core}}

\begin{figure}
\epsscale{.9}
\plotone{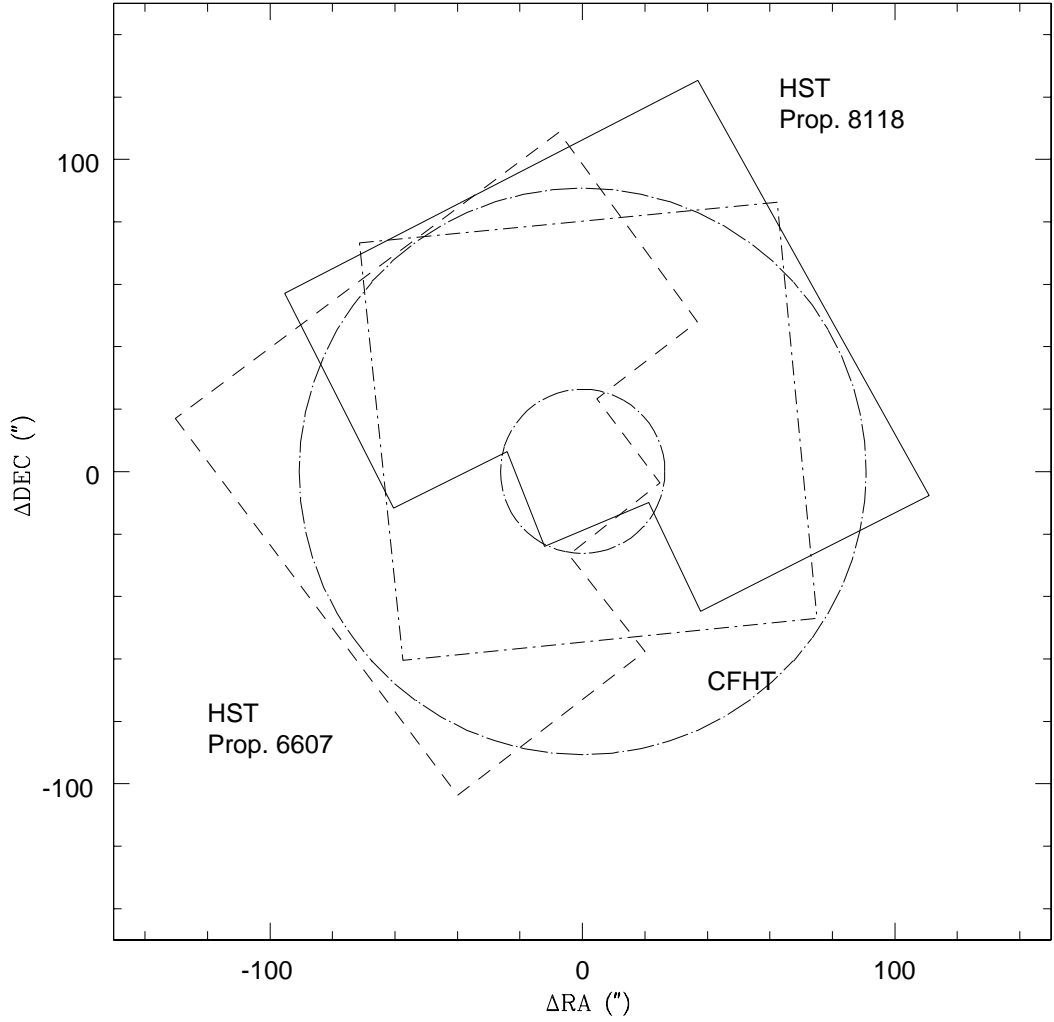}
\caption{Fields observed in the core of M5.  The inner circle represents the
core radius ($r_c = 26\farcs3$) and the outer circle represents the half-mass
radius ($r_h = 90\farcs75$).\label{fields}}
\end{figure}

Photometric data for the stars in the cluster core were matched by sky
position. Figure \ref{fields} shows the regions covered by the three datasets.
Because these fields did not completely overlap and different filter
combinations were used for each field, the brightness and color information
available varied from star to star. We used selection criteria similar to
those used by \citet{fer97} in their study of M3. BSSs were selected from the
four CMDs shown in Fig. \ref{corecmds}. The ($V, B-V$) and ($m_{255},
m_{255}-m_{336}$) HST CMDs were the primary choice for BSS selection due to
the high spatial resolution.  We defined the most reliable ``bright'' BSS
(BBSS) sample as stars with $B < 17.81$ or $m_{255} < 18.35$ (see \S
\ref{comp} for the motivation behind these cuts).  The ($m_{555},
m_{336}-m_{555}$) CMD was also used in the selection, but mainly as a
secondary check for portions of the field that did not overlap with the ($V,
B-V$) CMD.  We identified a star as a BSS if it was selected in both the ($V,
B-V$) and ($m_{255}, m_{255}-m_{336}$) CMDs if the star was covered by both
HST fields, or in just one if in a part of the field where the HST
observations did not overlap.

Reliable faint BSSs (FBSSs) were selected differently for each CMD.  For the
($V, B-V$) CMD, FBSSs were selected if they were significantly bluer
($B-V=0.37$) and brighter ($V=18.57$; \citealt{sand96}) than the MSTO and
below the BBSS cutoff. The color cut was chosen to avoid photometric scatter
near the turnoff. FBSSs were selected in the ($m_{255}, m_{255}-m_{336}$) CMD
by choosing all of the stars brighter than the majority of the photometric
scatter ($m_{255}<18.9$) and below the BBSS cutoff, though this is likely to
be contaminated by optical blends to some degree.  Stragglers in the
($m_{555}, m_{336}-m_{555}$) CMD were selected if they were brighter than the
MSTO and outside of the scatter. A faint limit cut of $m_{555}<18.31$ was
imposed.  We emphasize, however, that this faint sample was {\it not} used in
the main comparisons later in the paper because it is prone to contamination
resulting from stars with large photometric errors and optically-blended
stars.

Bright or faint BSSs were only selected from the CFHT data ($B, B-I$) if no
data were available from HST photometry because blending effects were
noticeable for stars nearest the cluster center. For the ($B, B-I$) CMD, FBSSs
were selected if they were bluer ($B-I=1.0$) and brighter ($B=19.04$;
\citealt{sand96}) than the MSTO and below the BBSS cutoff. The candidate FBSSs 
closest to the MSTO appeared to be within the scatter and were not
included, although FBSSs could exist in that region.  Once again, we emphasize
our purpose in tabulating FBSSs is simply to identify candidates for
future work --- we have not included them in the main analysis in this paper.

Stars that were not selected as BSSs that appear in the CMDs in regions that
BSSs occupy were eliminated based on information from other overlapping
photometric datasets. In total, 48 bright BSSs and and 47 faint BSSs were
selected.

\subsection{The Outer Region Sample}

The outer region of M5 consists of two ground-based data sets from CTIO: one
in $B$ and $V$ and the other in $B$ and $I$. Stars were again matched by
position, and put on a common coordinate system with the core stars. The
regions covered by the HST and CFHT core fields were then removed from
consideration.  Due to photometric scatter in the ground-based data, we only
selected stars that were above the BBSS magnitude cutoff ($B<17.81$) used in
the core sample.  The bright BSSs were then selected in the same manner as
the core sample.  A star was considered a BBSS if
$B<17.81$ and $B-I<1.05$. 

 As before, stars that appear in the ($V, B-V$) CMD that were not selected
    as BBSSs were rejected based upon
    their positions in the ($B, B-I$) CMD. The $B$ and $I$ images were taken
    under better seeing conditions than the $V$ images, and had fewer issues
    with blended star images. Based on these selections, we
    identified 11 BBSSs outside of our core sample.  
    
    Faint BSSs ($B > 17.81$) were selected after an additional radial cut was
implemented.  Stars just outside the {\it HST} and CFHT fields were still subject to
significant crowding effects and to minimize these effects
a radial cut ($r > 152\arcsec$) was implemented. The selection criteria are
discussed in more detail in \S\ref{gcomp}.  Only the ($B, B-I$) CMD
was used in the selection of faint candidate BSSs due to the noticable blending
effects seen in the ($V, B-V$) CMD even with the radius cut.  
A total of 13 FBSSs were selected.  The selected BBSSs and FBSSs are shown in 
Fig. \ref{g1cmd}. Fig. \ref{bsspos} shows the
cluster position of the BBSSs from the core and outer region samples. We
emphasize again that the FBSSs were not used in the analysis below, and are
included for completeness.

\section{Comparisons with Other Studies \label{comp}}

To make fair comparisons with studies of stragglers in other clusters, we have
taken the BSSs derived from all the photometry we had available (as described
above) and applied the photometric cuts described in those studies.  We have
also applied the same photometric cuts to samples from single CMDs to try to
choose BSSs in a manner most accurately matching that of the original paper.
In this way, we can minimize systematic errors and gauge possible pitfalls
involved in selecting stragglers using information from a single CMD.

\subsection{Comparison with \citet{fer97} \label{gcomp}}

\citet{fer97} examined M3 using multiple wavelengths [filters F255W, F336W,
F555W ($V$), and F814W ($I$) in the core, and $B$, $V$, and $I$ in the
outskirts] in order to obtain a global BSS sample for the cluster. They used a
cut in $m_{255}$ to select their core BSS sample, and a cut in $B$ to select
their outer sample.  \citeauthor{fer97} used the $m_{255}-m_{336}$ color to
select their core BSS sample because the area of the CMD occupied by BSSs is
very distinct from the locuses of single stars in other evolutionary phases.
They divided their sample into two sub-samples: BBSSs and FBSSs.  BBSSs were
selected in M3 if $m_{255} > 19.0$, while FBSSs had $19.0 < m_{255} < 19.4$
(though FBSSs were not used in their analysis).  The BSS sample was split in
order to avoid contamination arising from stars with large photmetric error or
optical blends. Later in this paper, we will compare the radial distribution
of BBSSs obtained for M3 by \citet{fer97} with the BBSS radial distribution
from M5. Here we first examine the consistency of these cuts because we have
HST photometry in $m_{255}$ and $B$ (more precisely, F439W calibrated to $B$)
in a portion of the core field.
 
To reliably compare our core sample with that of M3, we needed to determine
BBSS sample cuts that were compatible with those used by \citet{fer97}. To do
this, we made our photometry cuts at the same position {\it relative to the
cluster turnoff}. M3 has a MSTO magnitude of $B=19.54$ \citep{fer97.2} and
\citet{fer97} used a faint cut for the BBSS sample at $B = 18.6$.  We used
this difference of 0.94 mag to define the faint end of the sample, using the
MSTO value for M5 ($B=19.04$; \citealt{sand96}) to define the cut at $B=18.1$.
In order to select BBSSs in the ($m_{255}, m_{255}-m_{336}$) CMD we followed
the method used in \citep{fer03} (see \S \ref{bcomp} for details).  BBSSs were
not directly chosen from the ($m_{555}, m_{336}-m_{555}$) CMD, which covered
the same field, although it was used as a check. 

Because some stars were observed in both HST pointings, we were able to check
the consistency of the two faint cuts. We found that ten stars fell below the
selection line in the ($m_{255}, m_{255}-m_{336}$) CMD though they satisfied
the requirements for a BBSS in the ($V, B-V$) CMD (see the right panels of
Fig. \ref{bad}).  Two stars that were selected based on the ($m_{255},
m_{255}-m_{336}$) CMD would not have been selected in the ($V, B-V$) because
they were redder than the turnoff.  Nevertheless, if we keep all stars that were 
selected using {\it either} the $B$ or $m_{255}$ cuts following \citet{fer97} and
\citet{fer03}, we end up with 59 BBSSs from the core sample.

The inconsistency in the samples derived using cuts in different filters
indicates that the faint cut needs to be adjusted for one of the filters.  In
order to have as consistent a sample as possible, we shifted the BBSS cut
brightward in $B$. A BBSS cut of $B=17.81$ improves the agreement between the
samples in ($V, B-V$) and ($m_{255}, m_{255}-m_{336}$). The one star below the
BBSS cut line in the ($m_{255}, m_{255}-m_{336}$) CMD was kept in because of
its position among the other BBSSs in the ($m_{555}, m_{336}-m_{555}$) CMD.
The two stars that fell just below the BBSS cut line in the ($V, B-V$) CMD
were kept in due to their positions in the ($m_{255}, m_{255}-m_{336}$) CMD.
This choice gives us a sample of 48 BBSSs from the core of M5, as shown in the
left panels of Fig. \ref{bad}.

The region outside of the core samples was scrutinized in the same
    fashion.  Using our BBSS cut of $B=17.81$, we identified 11 BBSSs. (Once
    again, only the BBSSs were used for comparisons.) If we instead use a
    fainter cut in $B$ consistent with that of \citet{fer97}, we obtain 23
    BBSSs.

\subsection{Comparison with \citet{fer03} \label{bcomp}}

In another study by \citet{fer03}, BSSs were identified in the cores of five
globular clusters using HST data in the $F255W$ and $F336W$ filters, and were
compared to results for M3 \citep{fer97} (see \S\ref{gcomp}). Photometry of M5
was taken as part of the same program (program ID 6607, PI: Ferraro), but has
not been presented in the literature.

In order to make the comparison, we shifted our ($m_{255}, m_{255}-m_{336}$)
CMD in magnitude and color to overlie the brightest HB stars in the M3 CMD,
following the method used by \citet{fer03}. The shift in the ultraviolet
wavelength F255W was done by lining up the brightest portion of the HB of M5
with that of M3.  The straight line cut at magnitude $m_{255} = 19.0$ in F255W
for M3 was then used as the dividing line for our BBSS sample.  This magnitude
cut falls at $m_{255} = 18.35$ for M5.

\begin{figure}
\epsscale{.9}
\plotone{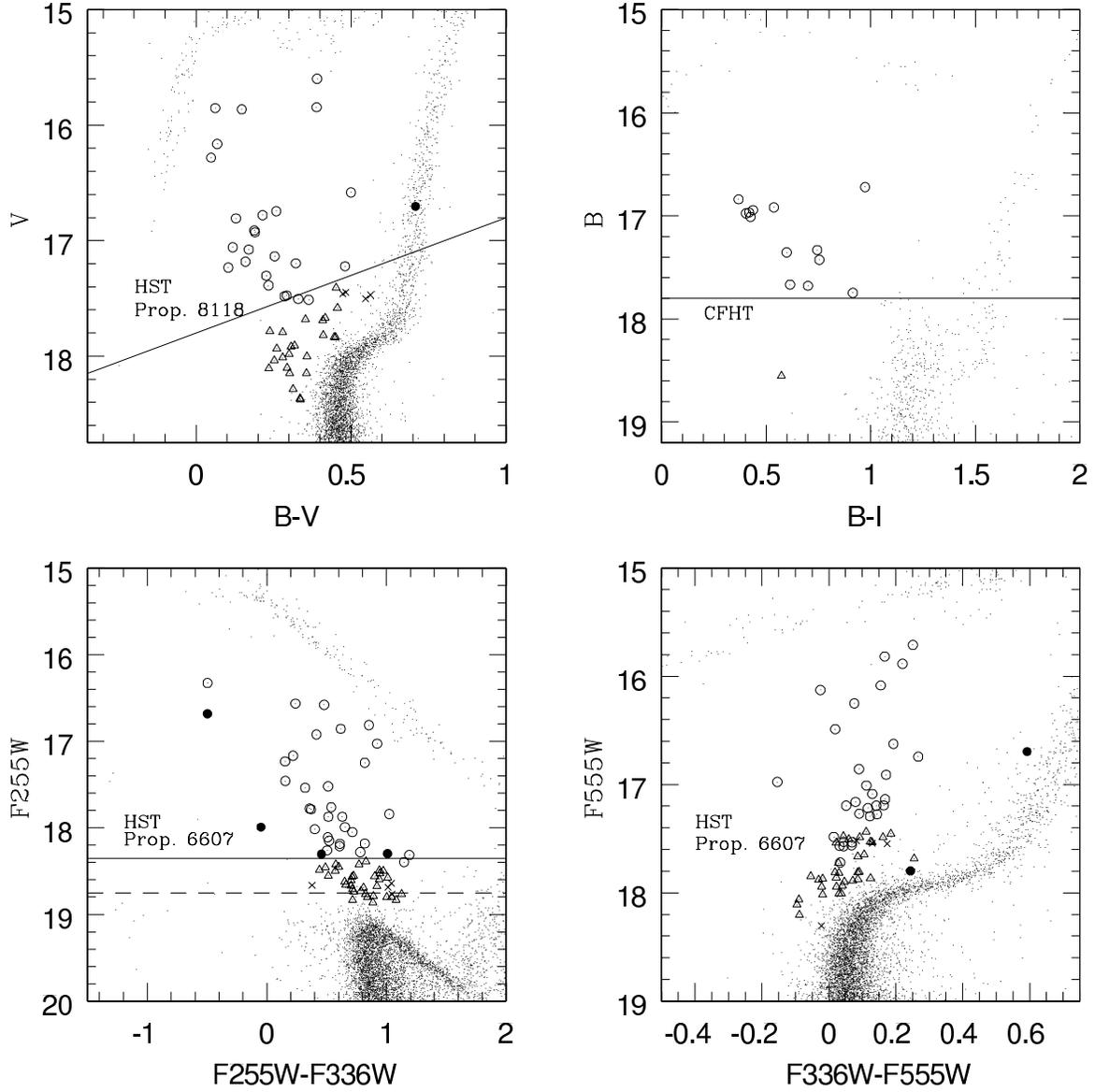}
\caption{CMDs for the core datasets. BBSSs are shown as
  circles and FBSSs are shown as open triangles. Crosses mark FBSSs
  with $B-V>0.47$. Filled black circles are BBSSs that {\it would} have been
  selected if {\it only} the information from the ($m_{255}, m_{255}-m_{336}$)
  CMD was used (see \S\ref{bcomp}).  The solid line in the ($m_{255}, m_{255}-m_{336}$) CMD
  corresponds to the BBSS cutoff at $m_{255}=18.35$ and the dashed line
  corresponds to the FBSS cutoff used by \citet{fer97} at $m_{255}=18.5$. The
  solid lines in the ($V, B-V$) and ($B, B-I$) CMDs represent the BBSS cut at
  $B=17.81$. Stars in the CFHT panel are only those falling outside the HST
  fields.\label{corecmds}}
\end{figure}

\begin{figure}
\epsscale{.9}
\plotone{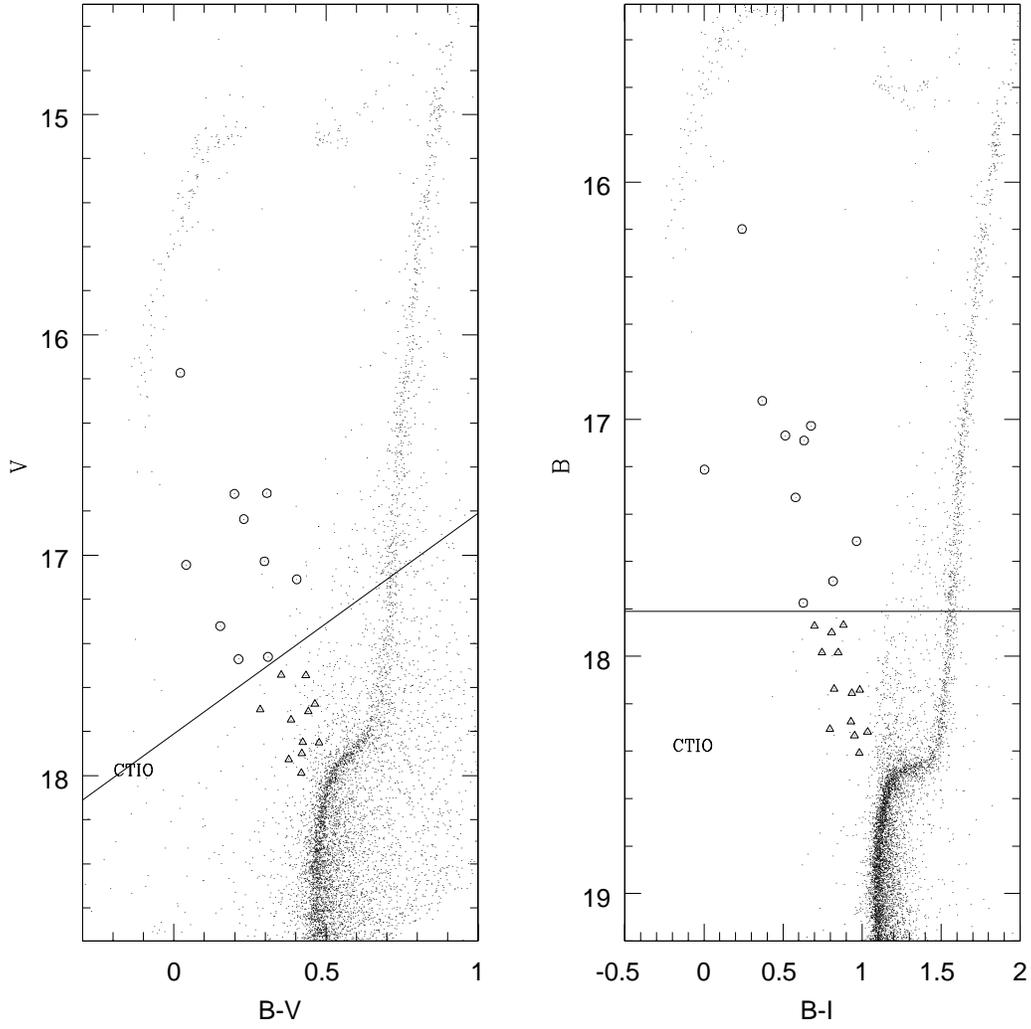}
\caption{Ground-based CMDs for the area of the cluster not covered by the HST and CFHT fields. 
BBSSs are circled and FBSSs are triangles.\label{g1cmd}}
\end{figure}

\begin{figure}
\epsscale{.9} 
\plotone{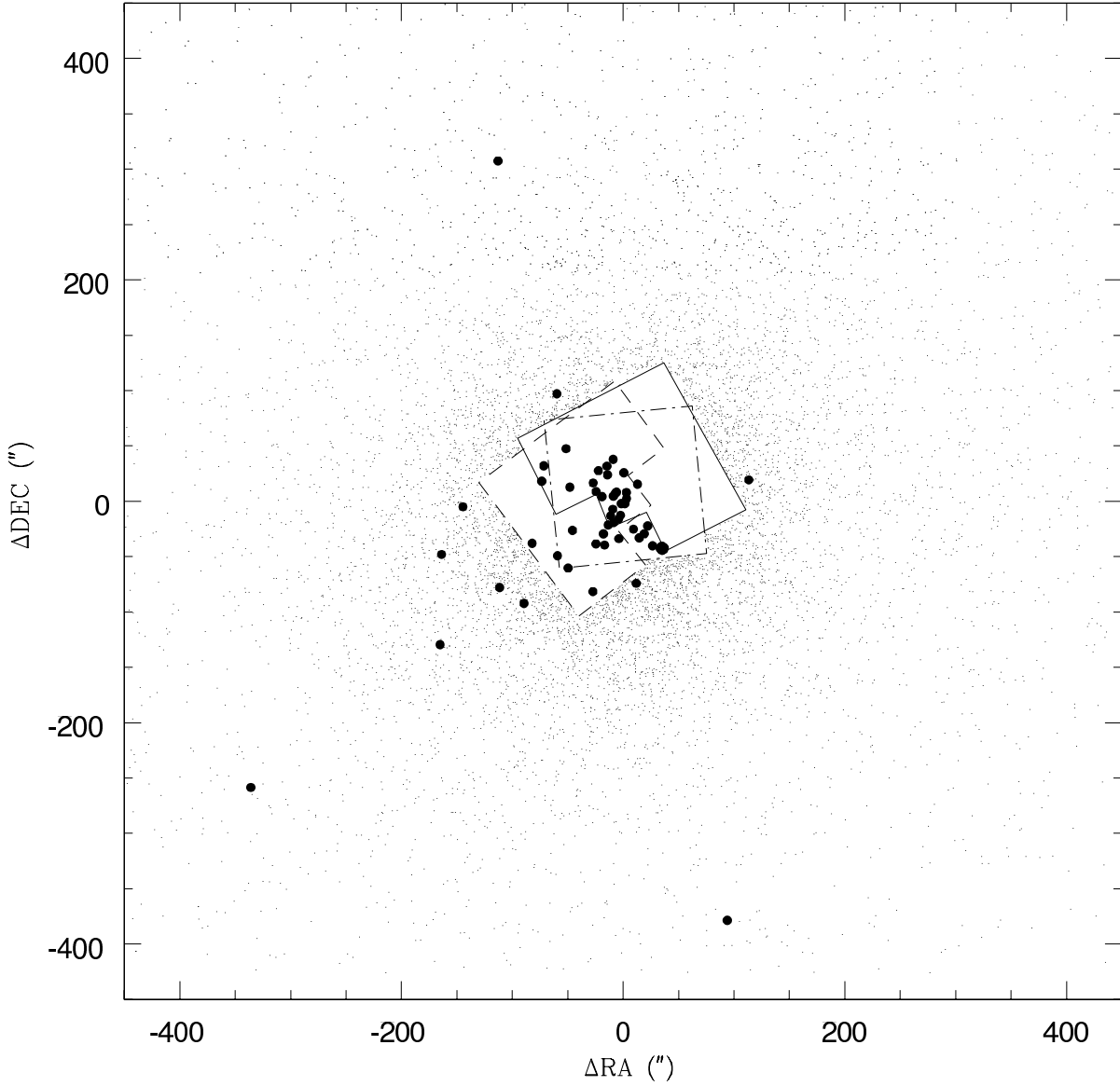}
\caption{BBSSs using our limit cuts are large black circles.  The non-BBSSs have been removed from
the core for clarity.\label{bsspos}}
\end{figure}

\begin{figure}
\epsscale{.9} 
\plotone{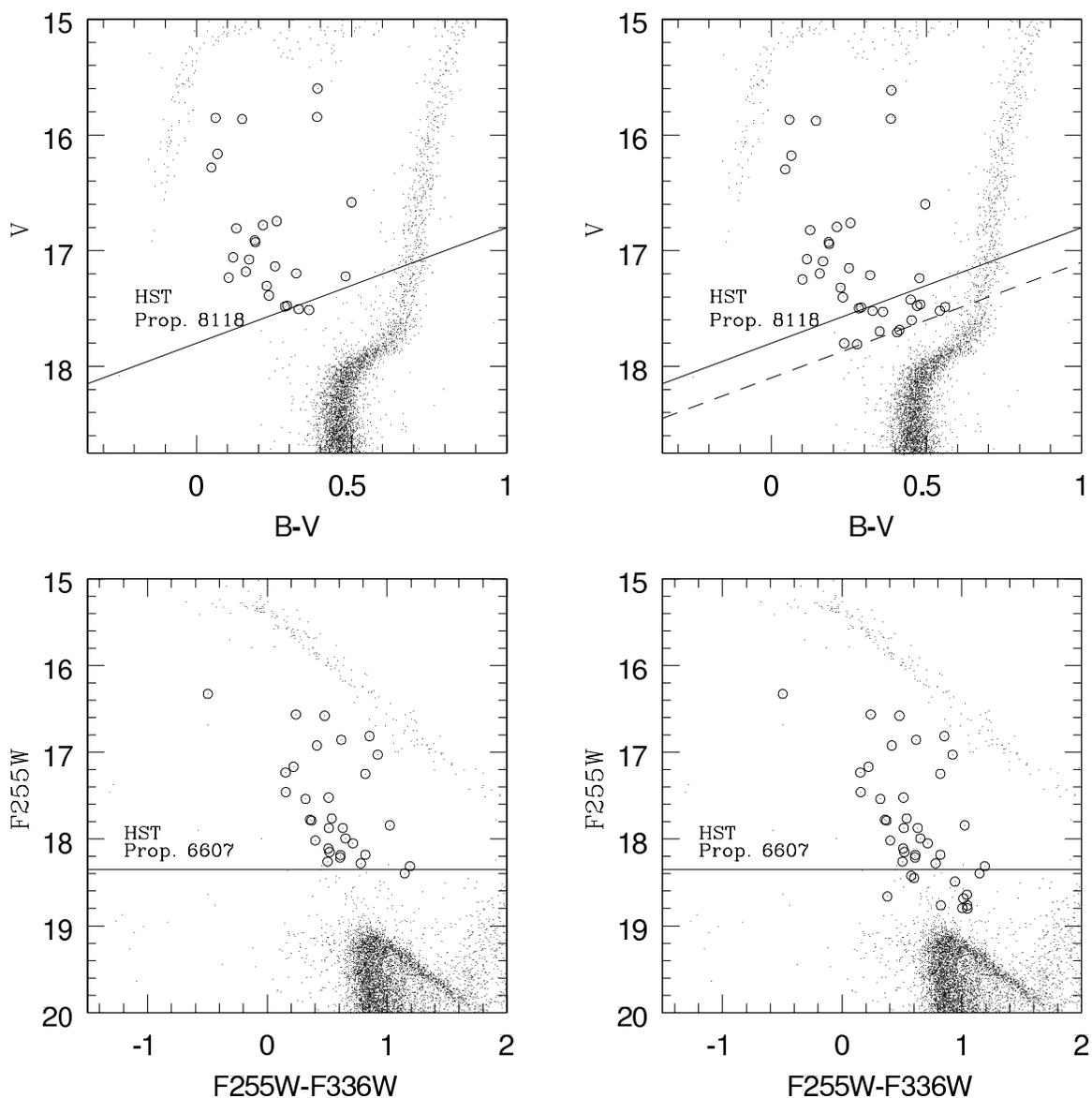}
\caption{The CMDs on the left show the BBSS sample selected using the cut
  (solid line) proposed by us at $B=17.81$, while the CMDs on the right show
  the BBSS sample selected using the cut (dashed line) from
  \citet{fer97} at $B=18.1$. We find better sample agreement in the upper and
  lower CMDs when the $B=17.81$
  cut is used. \label{bad}}
\end{figure}

Using our best BSS sample (selected using information from other CMDs when
available), we obtain a count of 31 BBSSs.  Fig. \ref{corecmds} shows the 31
BBSSs obtained from our BSS list. The one star below the BBSS cutoff was kept
because of its placement in the ($m_{555}, m_{336}-m_{555}$) CMD. If we use
{\it only} information from the ($m_{255}, m_{255}-m_{336}$) CMD to select
stars, the sample size is 34 BBSSs.  

The filled black circles in figure \ref{corecmds} are the extra BBSSs
    that would have been selected if we had used {\it only} the information
    from the ($m_{255}, m_{255}-m_{336}$) CMD.  In other CMDs, these stars
    were found far outside the regions where BSSs were found. The bluest of
    these stars in the ($m_{255}, m_{255}-m_{336}$) CMD of figure
    \ref{corecmds} was not found in the ($V, B-V$) CMD, but had a
    $m_{336}-m_{555}$ color of $-1.862$.  The next bluest star had a $V$
    magnitude of $18.4963$ and a $B-V$ color of 1.4198, and fell just above
    the SGB in the ($m_{555}, m_{336}-m_{555}$) CMD. The next star (moving
    redward in $m_{255}-m_{336}$) did not appear in the ($V, B-V$) CMD, but
    had $m_{336}-m_{555} =$ 0.993, which put it redward of the RGB in the
    ($m_{555}, m_{336}-m_{555}$) CMD. The reddest of the four stars in
    $m_{255}-m_{336}$ fell within the RGB in the ($V, B-V$) CMD and just
    blueward of the RGB in the ($m_{555}, m_{336}-m_{555}$) CMD.

\subsection{Comparison with \citet{pio04}}

\begin{figure}
\epsscale{.9} 
\plotone{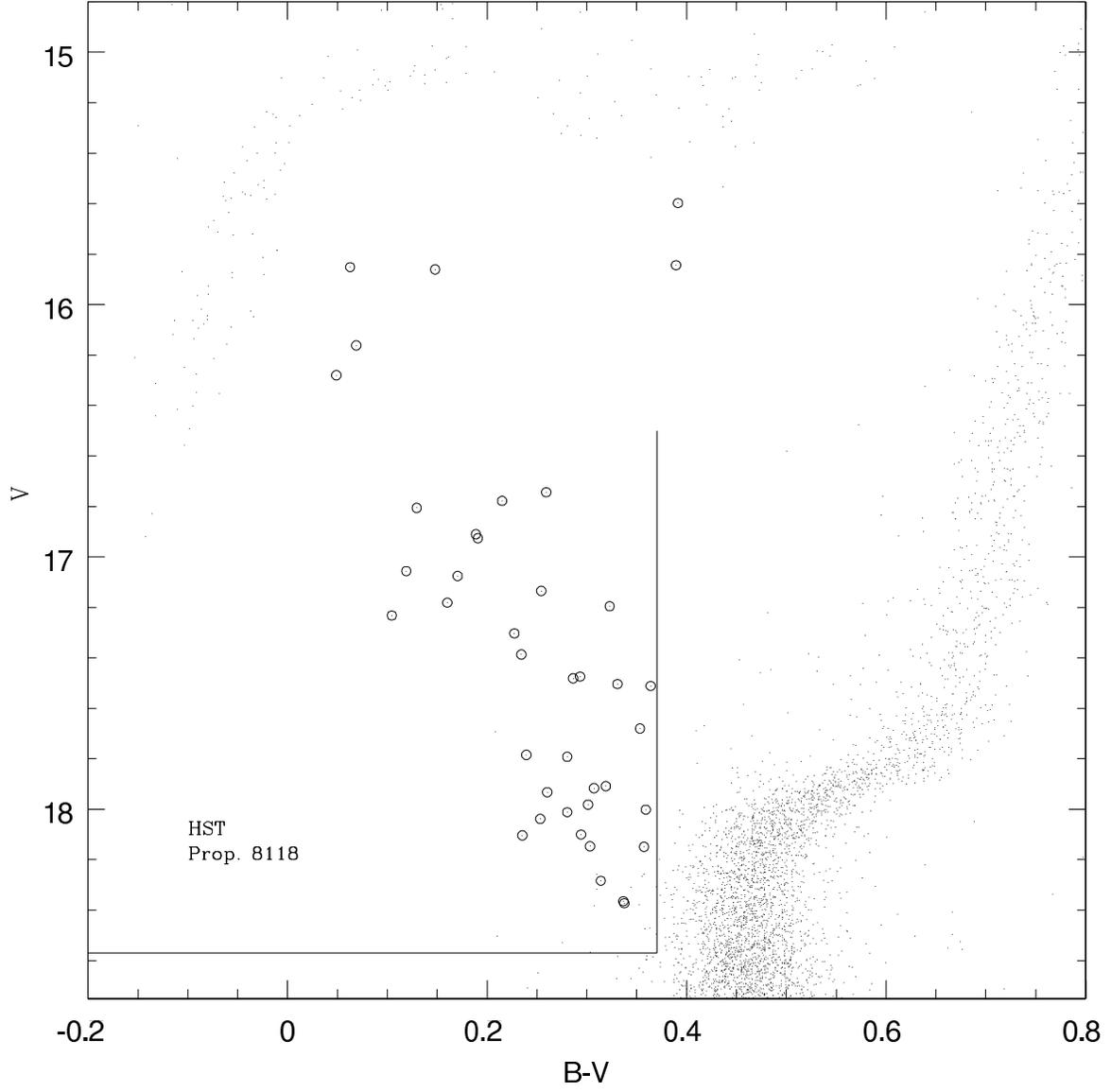}
\caption{BSSs used for the \citet{pio04} comparison are circled.\label{piocomp}}
\end{figure}

\citeauthor{pio04} selected blue stragglers in the cores of 56 globular
clusters (including M5) using $BV$ data from HST. We used the same images
(program ID 8118, PI: Piotto) in selecting a portion of our core sample for
comparison. \citeauthor{pio04} selected stragglers brighter than the MSTO of
each cluster, and for M5, the MSTO has a $V = 18.57$ and $B-V = 0.47$
\citep{sand96}. We chose all BSS (selected with information from
other CMDs when available) with $V<18.57$ and $B-V<0.37$. Our color cut was
chosen to avoid photometric scatter near the main sequence. The sample of 40
stars for this comparison can be seen in Fig.  \ref{piocomp}.  Again, stars
within this region of the CMD that were not circled were rejected using
information from other CMDs. Using the raw ($V, B-V$) CMD and the above
selection boundaries, we obtain a sample of 55 stars.

\section{HB Stars and the Brightest Stragglers}

To examine the frequency of stragglers, we used the HB and RR Lyrae stars
tabulated previously by \citet{sand04} as a comparison population. Six stars
in our core sample of BSS were identified as probable blends of HB and RGB
stars by \citeauthor{sand04} based on their position redder than the blue tail
of the HB.  (These stars had HB identification numbers of 107, 201, 221, 250,
268, and 276.) Using ultraviolet data from HST (with its better temperature
sensitivity for the hottest HB stars), we find that they are significantly
fainter than the HB locus, which makes us confident that they are bright BSS,
and not blends of HB stars.  One star in our outer region sample (HB ID number
302) was also reclassified as a BSS based upon its position redder than the
blue tail of the HB in both of the ground-based CMDs. No known RR Lyrae stars
made it into our BSS selections.

The BSSs that were mistakenly identified as HB stars by \citet{sand04} are an
interesting class of BBSSs. The most likely formation mechanism for these
stars is stellar collisions for the simple reason that
a merger of a close binary star or mass transfer within a binary cannot
produce a remnant that is massive and luminous enough. The
merger of at least three stars (resulting from a collision involving a binary
and at least one other star) seems to be the only feasible means of
accomplishing this.

Finally, we have three stars to add to the list of HB stars in \citet{sand04}
(see Table \ref{hbtab}).
Star 1 has a proper motion estimate \citep{res93} and was identified
as a likely non-member; however, it was the faintest star in the study and its
membership probability is quite uncertain.  It falls near the middle of the
blue tail of the HB in our ground-based $BV$ dataset, and so we consider it
likely to be an HB star. Star 2 was mistakenly excluded by
\citet{sand04} due to a bad $V$ band measurement that prevented it from
being identified as an HB star.
However, based upon its position in both of our CTIO CMDs near the tip of
the blue tail of the HB, it should be included. Star 3 is close to a
bright giant star and its photometry may have been contaminated.  Its CMD
position in the $BV$ HST data (brighter and bluer than the HB) leads us to
believe it is an evolved HB star. 

\section{Results}

\subsection{The BBSS Radial Distribution\label{raddist}}

\begin{figure}
\epsscale{.9}
\plotone{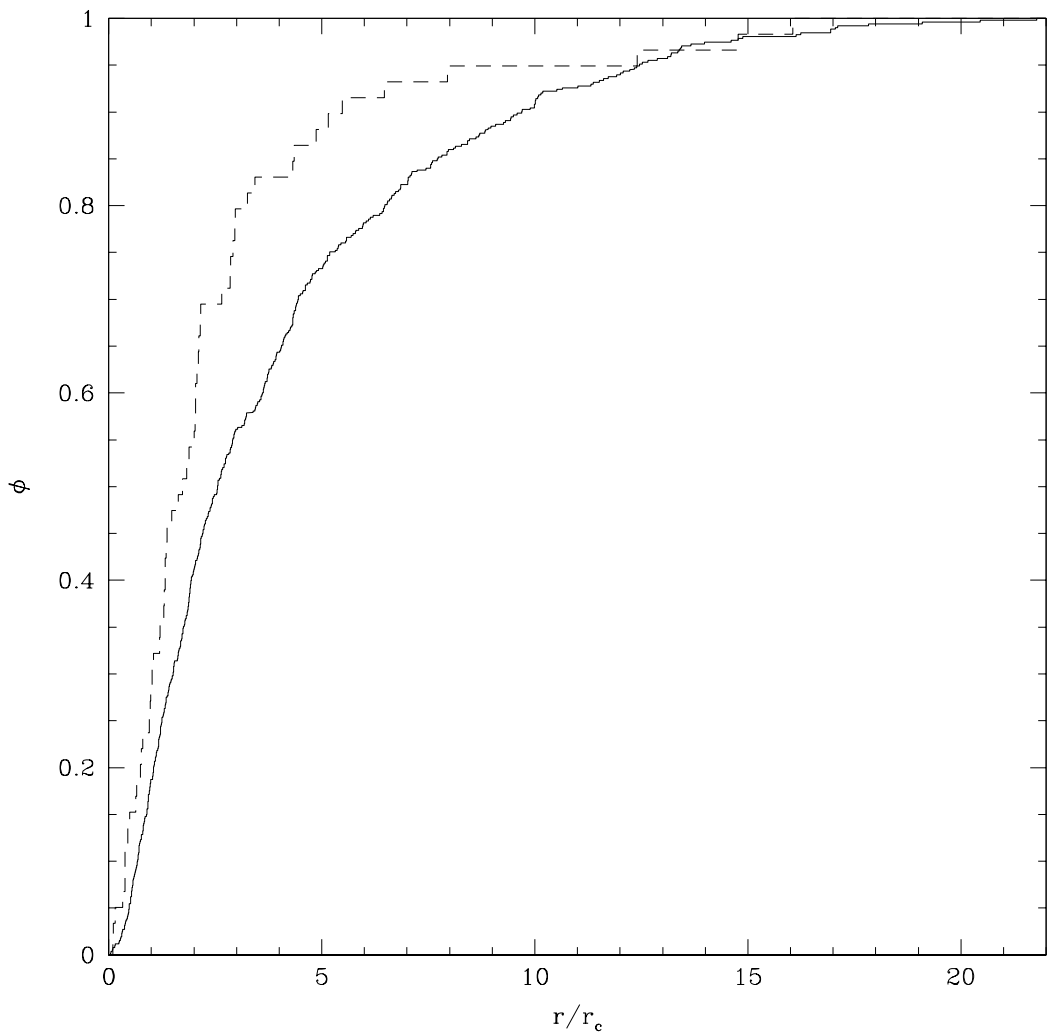}
\caption{The cumulative radial distributions for the BBSSs (dashed line) and
the HB stars (solid line) of M5.
\label{hbkstest}}
\end{figure}

 In most clusters in which stragglers have been surveyed, the stragglers
  have been found to be more centrally concentrated than comparison
  populations (such as HB or RGB stars). For example, \citet{fer97} showed
  that the BBSSs of M3 were more centrally concentrated than the cluster's RGB
  stars. In Fig.  \ref{hbkstest}, we compare the cumulative radial
  distributions of the 59 total BBSSs\footnote{This sample is composed of our
    samples of 48 core stragglers and 11 envelope stragglers using the most
    consistent choice of faint cuts, as described in \S \ref{gcomp}.} and the
  513 HB stars in the same field \citep{sand04}.  The figure shows that the
  BBSSs are {\it more} centrally concentrated than the HB stars, and a Kolmogorov-Smirnov (K-S)
  test gives a probability of 0.0015 that the two samples come from the same
  distribution.

For the globular cluster M3, \citeauthor{fer93} discovered a bimodal
distribution with a centrally-concentrated core population, a
sparsely-populated ``zone of avoidance'', and an outer population.
\citet{sig94} attempted to explain the origin of the separate BSS populations,
suggesting that all BSSs are formed via collisions involving primordial
binaries in the cluster core. The BSS that forms can be kicked into the
outskirts of the cluster if the ``spectator'' star (in binary-single star
collision) is ejected from the system or if there is sufficient mass loss
during the collision to perturb the center of mass. \citealt{sig94} also
suggest that BSSs formed or kicked out to $\lesssim 7 r_{c}$ will quickly
migrate back into the core via dynamical friction, while those kicked out
beyond $\sim 7 r_{c}$ are effectively ``parked'' in the cluster outskirts.
The two different timescales will naturally give rise to the ``zone of
avoidance'' seen in M3.

Since the study by \citet{fer93}, similar distributions have been confirmed in
a few other massive clusters (M55, \citealt{zag}; 47 Tuc, \citealt{fer04}; NGC
6752, \citealt{sabbi}), while a unimodal distribution was observed in one
low-mass cluster (Palomar 13, \citealt{clark}). Because M5 has a structure
that is quite similar to that of M3, we can test whether structurally-similar
clusters produce similar straggler distributions.  According to the values in
Table \ref{m3m5} \citep{har96,mclaugh}, M3 and M5 have similar concentrations
$c$, while M5 has a 17\% smaller core radius $r_c$ than M3, and slightly
smaller half-light radius $r_h$\footnote{The half-light radius is more
difficult to determine than core radius because it depends on 
accurate knowledge of the whole surface brightness profile. We have quoted
values from the Wilson models presented in \citet{mclaugh} because they
follow the distribution at large distance from the cluster center more closely
than King models do. Based on this, the half-light radius of M5 is about 11\%
smaller than that of M3. When King models are used instead, M3's half light
radius becomes considerably larger, so that M5 appears to have a 45\% smaller
half-light radius.}.

\begin{figure}
\epsscale{.9}
\plotone{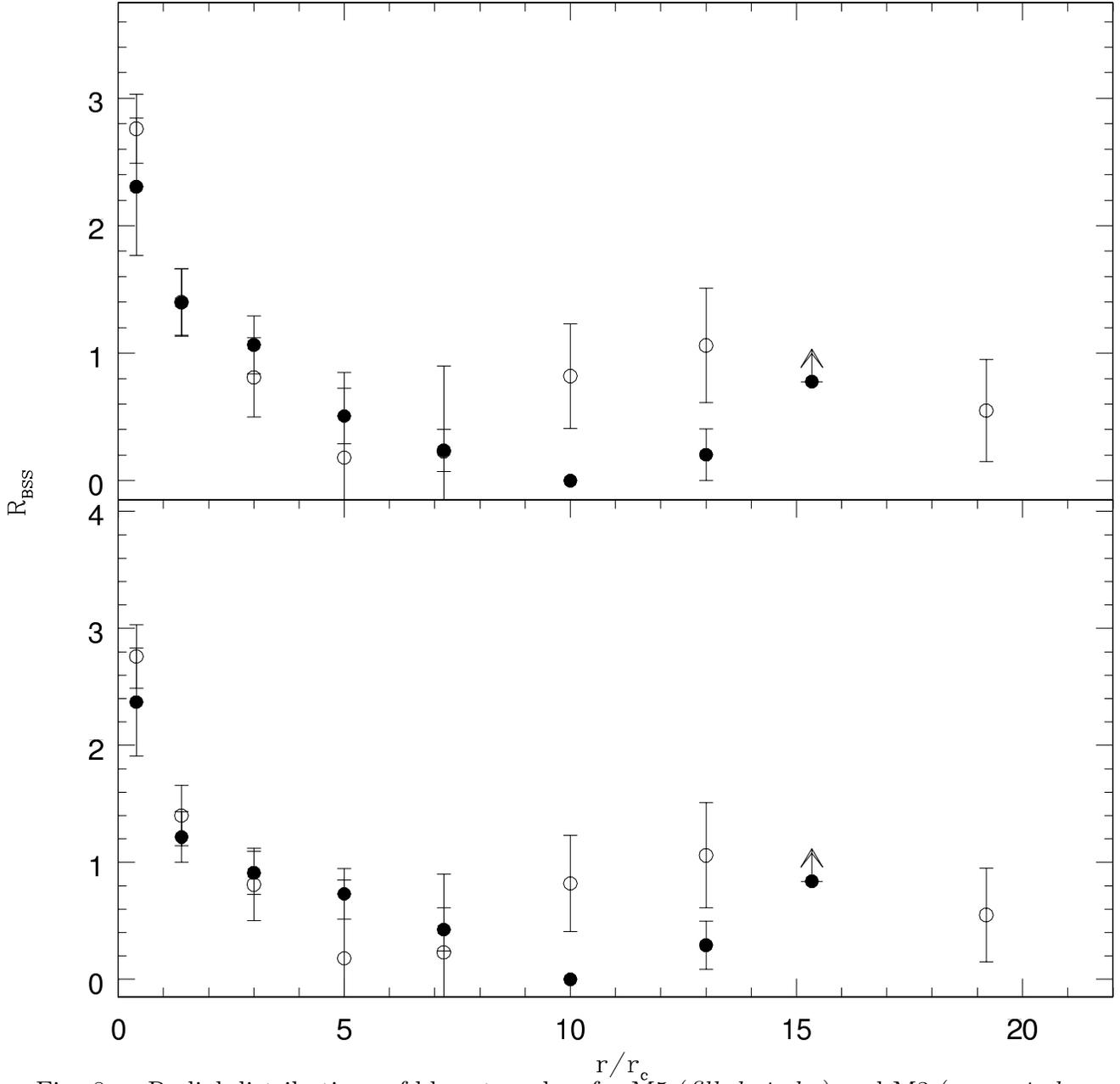}
\caption{Radial distributions of blue stragglers for M5 ({\it filled circles})
  and M3 ({\it open circles}; \citealt{fer97}). The top panel shows the radial
  distribution for M5 when our BBSS cuts are used (see \S\ref{gcomp}), while
  the bottom plot shows the radial distribution when the BBSS cuts of
  \citet{fer97} are used.\label{rbss}}
\end{figure}

In Fig. \ref{rbss}, we compare the radial distribution of our BBSS sample with
that of \citet{fer97} for M3 when scaled horizontally for differences in core
radius. The quantity $R_{BSS}$ \citep{fer93} is
\[
R_{BSS} = \frac{(N_{BSS}/N_{BSS}^{tot})}{(L^{sample}/L_{tot}^{sample})} .
\]
The radial distribution for M5, like that for M3, has a bimodal nature,
although our field was not large enough to identify where the outer population
of BSS peaks. Scaling by core radius is a natural choice if the distribution
is modified by mass segregation of stragglers through dynamical interactions
with other cluster stars. This scaling does indeed cause the distributions to
overlie each other in the central regions of the clusters. Nonetheless, we
examined alternate ways of scaling the radial distributions to compensate for
the small structural differences between clusters. Scaling by half-light
radius introduces more uncertainty because the half-light radii are harder to
determine (see the footnote in \S \ref{raddist}).  In the case of M3 and M5
though, scaling by half-light radius would be very similar to scaling by core radius.  We also tried using $L^{sample}/L_{tot}$
(derived from King models) as a horizontal coordinate, although this again did
not produce an alignment of the straggler distributions. \citet{sabbi} provide
additional evidence for a scaling with core radius in their examination of
straggler populations in the cores of 8 clusters from HST data, finding that
the half-sample radii fell at approximately 1 $r_c$. (Their samples, however,
only included stars within single HST WFPC2 fields of view.)

When the two cluster distributions are scaled with core radius, we find that
they agree to within $2 \sigma$ out to $r/r_{c} \la 13$.  M5 appears to have a
wider ``zone of avoidance'' ($4 \la r/r_c \la 14$) than M3 ($4 \la r/r_c \la
8$), and a lower frequency of stragglers there. Both of these facts imply that
M5's structure is somewhat more dynamically evolved (stars that started
farther from M5's center having been dynamically relaxed to the core).

\begin{figure}
\epsscale{.9}
\plotone{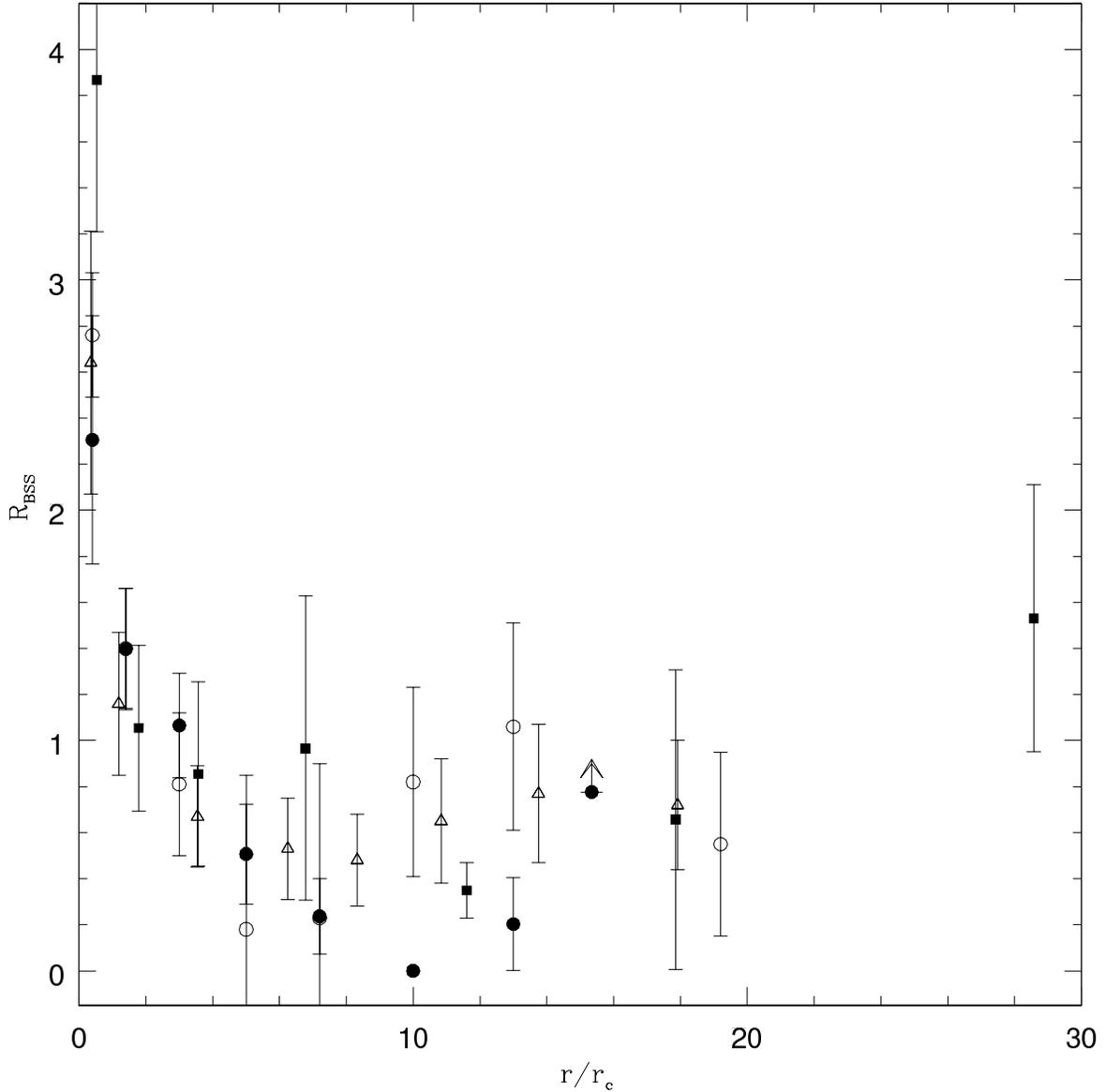}
\caption{Radial distributions for M5, M3, 47 Tuc, and NGC 6752. Filled
circles are data for M5, open circles are data taken from
\citet{fer97} for M3, open triangles are data taken from \citet{fer04}
for 47 Tuc, and filled squares are data taken from \citet{sabbi} for
NGC 6752.\label{rbss2}}
\end{figure}

BSS radial distributions have only been produced for a handful of other
clusters.  Figure \ref{rbss2} shows the scaled radial distributions for M5, M3
\citep{fer97}, 47 Tuc, and NGC 6752. (The distribution for NGC 6752 uses the
quantity $(N_{BSS} / N_{BSS}^{tot}) / (N_{HB} / N_{HB}^{tot})$ as a stand-in
for $R_{BSS}$.) All four clusters have a core BSS sample that falls within
about $4 r_c$ of the center.  The minima of the distributions all fall in the
range $5 \la r/r_{c} \la 12$.

\citet{map04} created dynamical models to understand the BSS distributions in
47 Tuc, finding it necessary to employ two separate populations of BSSs: a
core population and an outer population, compared to a single core population
described by \citet{sig94}. The region $r < 20r_{c}$ (containing the core and
current zone of avoidance) produced BSSs primarily through stellar collisions,
which were concentrated to the center of the cluster through the action of
dynamical friction.  The region beyond $20 r_c$ was populated with BSSs born in
primordial binary systems via mass transfer, and remaining in the cluster
outskirts due to the weakness of dynamical friction in the low density
environment. These ingredients can naturally produce zones of avoidance seen
in all four clusters in Fig.  \ref{rbss2}, although differences in cluster
history and current cluster characteristics could leave signatures on the
straggler distribution.

\citeauthor{map04} used the timescale for dynamical friction
\citep{bin87}
\[
t_{DF} = \frac{3}{4 \ln (\Lambda) G^{2}
(2\pi)^{1/2}}\frac{\sigma^{3}}{M\rho(r)}
\]
to predict the radius at which the zone of avoidance should occur, assuming
circular star orbits within the cluster, cluster ages of about 12 Gyr, typical
BSS mass $M$ of about $1.5 M_{\odot}$, Coulomb logarithm $\ln
\Lambda \approx 10$, and appropriate King models for the structure. Using
velocity dispersion values $\sigma$ given by \citet{pry93}, we obtained
expected $r/r_{c}$ values for M5 ($\thicksim8.1$), M3 ($\thicksim6.3$), 47 Tuc
($\thicksim8.3$), and NGC 6752 ($\thicksim14.6$) that agree nicely with the
distributions shown in Fig.  \ref{rbss2}. The assumption of circular orbits is
probably the most questionable --- if the straggler orbits are significantly
eccentric, this would shorten the dynamical friction timescale since the
stragglers would periodically sample higher stellar densities and higher rates
of dynamical friction.

\begin{figure}
\epsscale{.9}
\plotone{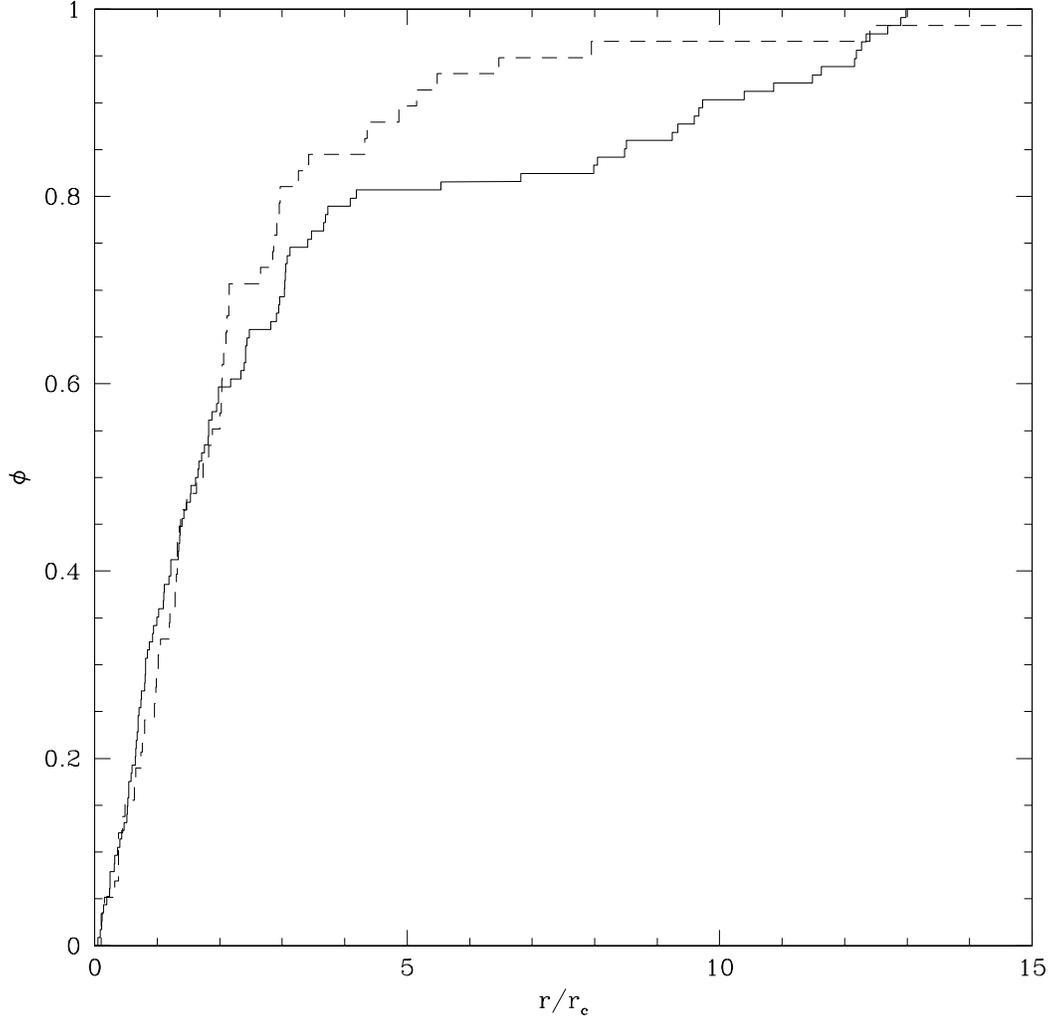}
\caption{The cumulative radial distributions for M3 ({\it solid line}) and M5
  ({\it dashed line}). The radial coordinates are scaled for the different 
core radii of the two clusters.
\label{kstest}}
\end{figure}

In Fig. \ref{kstest} we plot the cumulative radial distributions of the BBSSs
of M5 (dashed line) and M3 (solid line).  A K-S test
applied to the two lines gives a 40\% probablility that the two radial
distributions are derived from the same underlying distribution. The
inconclusive nature of the test results mostly from the relatively small
number of bright stragglers in M5 and the fact that the largest
  differences are in the sparsely populated outskirts of the clusters. Because
  K-S tests are not as sensitive to differences in the tails of distributions,
  we conducted Monte Carlo simulations by randomly drawing 58 BBSSs (the
  size of the M5 sample which spans the M3 sample, $r<360\arcsec$) from the M3 sample (114 stars).  
  We looked for trials
  in which the maximum deviation between the cumulative radial distributions
  was greater than 0.14 (the measured deviation between the M5 and M3
  samples), with the smaller sample having the larger cumulative distribution
  value. For ten million trials, we found that the frequency of this 
 was 9.5\%, which indicates that the M5 distribution is
marginally inconsistent with the M3 distribution.

The largest differences between the BBSS populations of M5 and M3 are in the
``zone of avoidance''. In making these comparisons, we must keep in mind that
the ordinates for the two distributions are not identical, and that these
tests do not account for the possibility of systematic errors in the
independent variable. For example, errors in the core radii would affect the
comparison of the distributions. In this paper, we used $r_c = 25\arcsec$ for
M3 from \citet{webb} as used in \citet{fer93} and subsequent papers. This
value differs slightly from some more recent fits to surface brightness data
($22\farcs9$, \citealt{mclaugh}; $30\farcs2$, \citealt{trager}). By
comparison, the most recent determinations for M5 agree very well ($26\farcs4$,
\citealt{lehm97}; $26\farcs3$, \citealt{mclaugh}; $24\arcsec$,
\citealt{trager}). Using the most recent value for M3 from the King model fit
of \citeauthor{mclaugh} (which agree well with values that can be derived by
interpolations in the surface brightness data), the differences between the M3
and M5 distributions become only slightly more significant (a K-S test returns
a 31\% probability that the M3 and M5 distributions come from the same
underlying distribution, and the Monte Carlo simulations indicate that stars
drawn from the M3 distribution produce M5-like distributions only 6.7\% of the
time).  More definitive conclusions will require larger straggler samples.

Regardless, King models scaled by core radius and by
central density agree very well for $r / r_c \la 10$ for a wide range of
concentrations $c$ (for example, Fig. 4-9 of Binney \& Tremaine 1987).  So if
populations of stars in two different clusters follow the underlying light
distributions (described by an appropriate King model), scaling by the core
radius should minimize differences in the scaled distributions.  The agreement
of the two blue straggler distributions for $r/r_{c} < 2$ provides some
supporting evidence for this choice.

\subsection{The Total BSS Population}

Apart from the radial distributions, the sizes of straggler populations vary
from cluster to cluster.  \citet{pio04} found that the relative frequency of
stragglers, defined as
\[
F_{BSS} = \frac{N_{BSS}}{N_{HB}} ,
\]
is significantly anti-correlated with total cluster luminosity.  However, the
scatter in the anti-correlation discussed by \citealt{pio04} is evidence of
some {\it additional} variation from cluster to cluster.  Using the comparison
sample obtained from our total BSS list for M5 (stragglers selected
using all available CMD information), we find $F_{BSS} = 0.239 \pm 0.042$. If
we select BSSs in a manner mimicking that of \citet{pio04}, we find $F_{BSS} =
0.329 \pm 0.051$. Both of these values are similar to the M5 value ($\sim
0.26$) used in Fig. 1 of \citet{pio04}. Regardless of the exact value, M5 (and
M3) fall near the upper envelope of clusters in the trend of log $F_{BSS}$
with $M_{V_t}$.

We then examined the global sample of bright stragglers for M3 and M5. The
global sample for M3 found by \citet{fer97} consisted of 114 BBSSs with $r <
360 \arcsec$. According to a King model for the cluster, this radius
corresponds to a fractional contained luminosity of 0.697.  To compare the
global sample for M3 to that of M5, we used a global sample selected with the
same criteria used by \citet{fer97} out to a radius having the same fractional
contained luminosity ($r < 244\arcsec$). As described earlier, we have also 
made a strong effort to ensure that our photometric selection criteria
matched the ones used for M3. Our sample contains 82 BBSSs,
returning a frequency $F_{BSS} = 0.159 \pm 0.019$ for M5, compared to $0.197
\pm 0.020$ for M3. Though M5 is the more luminous cluster, it contains about 
20\% fewer bright stragglers than M3.

Using a single HST WFPC2 field of view centered on the cores of 8 clusters,
\citet{sabbi} compared the specific frequencies of {\it bright} BSS defined as
\[
F_{HB}^{BBSS} = \frac{N_{BBSS}}{N_{HB}} .
\]
Using the 31 BBSSs found in our UV sample, we obtain a frequency of
$0.19\pm0.04$ for M5.  When the BBSSs are selected using the method of
\citet{fer97}, we find $F_{HB}^{BBSS} = 0.21\pm0.04$.  These two values for M5
are statistically consistent and give M5 one of the {\it lowest} values of the
clusters listed in Table 3 of \citet{sabbi}.  \citeauthor{sabbi} found that
NGC 6752 had a low specific frequency ($0.18\pm0.04$) of bright stragglers,
while \citet{fer97} found a frequency of $0.28\pm0.04$ for M3.  The specific
frequencies for M5
and M3 do not differ greatly, as was the case for the radial distributions.
This again shows that there is little difference between the two clusters' BSS
populations.  We must keep in mind that even among the clusters examined by
\citeauthor{sabbi} there was a wide range of cluster luminosities represented,
so that simple comparisons of specific frequencies could be misleading.

\begin{figure}
\epsscale{.9}
\plotone{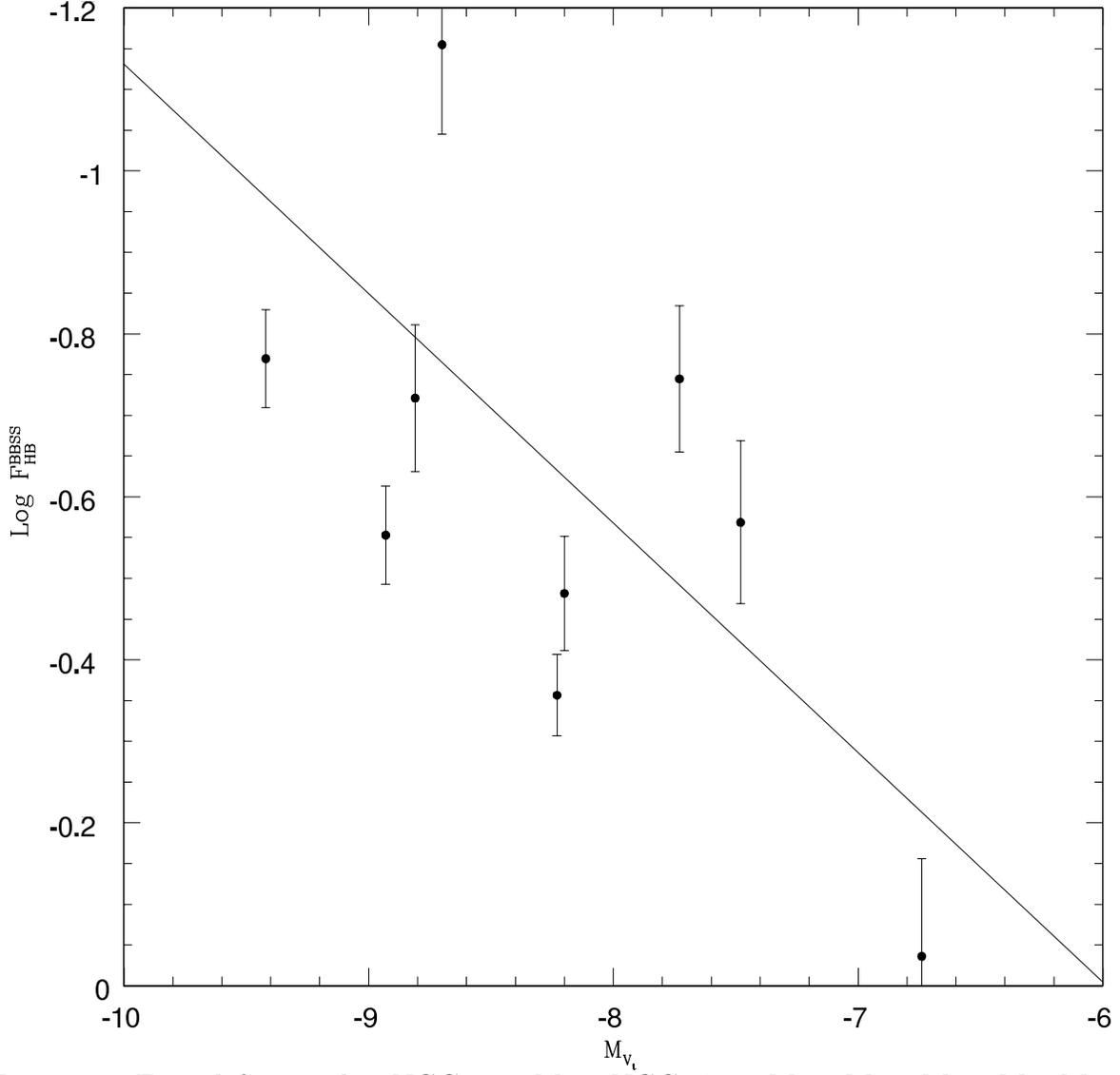}
\caption{From left to right:  NGC 288, M10, NGC 6752, M92, M80, M13, M5, M3, and
47 Tuc.  The best fit line through the data points has an equation $y = 0.28x + 
1.68$ with an error in the slope of $\pm0.11$.\label{spec}}
\end{figure}

If the brightest blue stragglers are more likely to be produced by collisional
processes (mergers of two or more stars), it is worth asking whether the
bright straggler frequencies show the same anti-correlation with cluster
luminosity that the larger samples of \citeauthor{pio04} do. As
\citeauthor{pio04} showed, there appears to be a change in the luminosity
function of stragglers between the brightest ($M_{V_t} \la -8.8$) and faintest
($-7 \ga M_{V_t} \ga -8.8$) clusters in their sample.  A least-squares fit to
the $\log F_{HB}^{BBSS}$ values ($M_{V_{t}}$ obtained from \citealt{har96})
(see Fig. \ref{spec}) for M5 and the 8 clusters studied by \citet{sabbi}
produces a slope (0.28) that is significant at the $2.5\sigma$ level.  This
slope is similar to the average slope seen in Fig. 2 of \citet{sand05} for
total samples in clusters with $-4 \ga M_{V_t} \ga -10$. So, with the limited
(in number of clusters and in luminosity range) cluster sample available at
present, the bright stragglers appear to be behaving similarly to the total
samples, although if we remove the lowest luminosity cluster (NGC 288) from
the list, the fitted slope is only $0.9\sigma$ different from zero.

The relative BBSS population sizes of M5 and M3 are very similar, as one might
expect for two clusters having similar dynamical properties.  Their frequency
values differ slightly, possibly in part due to the small differences in their
total cluster luminosities. The two clusters may differ in the distribution of
their stragglers.  M5's ``zone of avoidance'' appears to be wider than that of
M3 and there is a larger percentage of the stragglers in this zone in M3 than
in M5. Both of these facts are consistent with the idea that M5's stragglers
are more dynamically relaxed than those of M3,
separating the core and envelope stragglers to a greater degree.

\section{Conclusions}

As was described years ago by \citet{fer97}, multi-wavelength photometry is
truly necessary to reliably identify BSSs by utilizing cross-checks
between the different bands.  The BSSs mistakenly identified as HB stars by
\citet{sand04} are a prime example of this. For clusters with blue HB
morphologies, the identification of bright blue stragglers is not a trivial
matter. These BSSs are of special interest because the brightest ones are
likely to be binary star collision products.  The UV bands are of great
importance when picking out these BSSs because they can most cleanly separate
stragglers from the hotter HB stars.  Optical colors are simply not
color-sensitive enough for the hottest stars.

Our results from the comparisons of the BSS populations confirm that M5 and M3
have similar dynamical characteristics.  Both of their radial distributions
show bimodality, implying distinct core and envelope populations. When scaled
for differences in core radius, the two distributions largely overlie each
other --- the agreement is best in the core, though M5 appears to have a
larger ``zone of avoidance'' and a secondary frequency peak that is farther
from the cluster center. The BSS frequencies for the two clusters are quite
similar as well.  The slightly smaller frequency values for M5 are consistent
with M5's higher total luminosity when considering the anti-correlation
between frequency and cluster luminosity mapped out by \citet{pio04}.

While cluster luminosity seems to have been established as a primary predictor
of straggler frequency, there does appear to be significant scatter that
results from another source.  Future work to understand this scatter should
include an investigation into other clusters with luminosities comparable to
M3 and M5, such as M2, M13, M53, and NGC 5824. In the clusters examined by
\citeauthor{pio04}, there is approximately a factor of 3 difference in
frequencies among clusters with integrated luminosities similar to M5. Since
these clusters have large total straggler populations, it would also be
possible to examine the radial distibutions in maximum detail.

\acknowledgements

We would like to thank undergraduate Tom Jackson for his contributions to
a preliminary study of M5's stragglers. Jackson was a participant in 
the NASA College Summer Research Program at Northwestern University.
This work has been funded through grants AST 00-98696 and 05-07785 from
the National Science Foundation to E.L.S. and M.B.

\clearpage

\begin{deluxetable}{lcc}
\tablewidth{0pt} 
\tablecaption{Approximate Structural Parameters of M3 and M5}
\tablehead{\colhead{} & \colhead{M3} & \colhead{M5}}
\startdata
$M_{V_t}$\tablenotemark{a} & $-8.93$ & $-8.81$ \\
$\log j_0$ ($L_{\odot} \mbox{pc}^{-3}$)\tablenotemark{b} & 3.47 & 3.90 \\
$c$\tablenotemark{b} & 1.89 & 1.71 \\
$r_c$ (arcsec)\tablenotemark{b} & $22\farcs9$ & $26\farcs3$ \\
$r_c$ (pc) & 1.2 & 0.96 \\
$r_h$ (pc)\tablenotemark{b} & 3.7 & 3.3 \\
$d$ (kpc)\tablenotemark{ab} & 10.4 & 7.5\\
\enddata
\label{m3m5}
\tablenotetext{a}{\citet{har96}, updated February 2003}
\tablenotetext{b}{\citet{mclaugh}}
\end{deluxetable}

\begin{table}
\begin{center}
\caption{Candidate BSSs.  The top half are core BSSs and the bottom half are
BSSs from the outer region sample. \label{CBSS}}
\begin{tabular}{crrcrcrc}
\tableline\tableline
Star & $\Delta \alpha$ (arcsec)\tablenotemark{a} & $\Delta \delta$ (arcsec)\tablenotemark{a} & Filter\tablenotemark{b} & 
Magnitude & Filter\tablenotemark{b} & Magnitude & Telescope \\
\tableline
   1 &$0.689$ &$25.780$ &V &$15.5975$  &B &$15.9888$    &HST\\
   1 &$     0.689$ &$    25.780$         &B &$  16.0518$  &I &$  15.0094$   &CFHT\\
   1 &$     0.689$ &$    25.780$        &F5 &$  15.7100$ &F3 &$  15.9610$    &HST\\
       & &                     &F2 &$  16.8150$& &\\
   2 &$     1.534$ &$    -2.248$         &V &$  17.3030$  &B &$  17.5303$    &HST\\
   2 &$     1.534$ &$    -2.248$        &F5 &$  17.2750$ &F3 &$  17.4190$    &HST\\
       & &                     &F2 &$  17.7770$& &\\
   3 &$     1.647$ &$    18.375$         &V &$  18.1037$  &B &$  18.3390$    &HST\\
   3 &$     1.647$ &$    18.375$         &B &$  18.2611$  &I &$  17.3427$   &CFHT\\
   3 &$     1.647$ &$    18.375$        &F5 &$  18.1090$ &F3 &$  18.0140$    &HST\\
       & &                     &F2 &$  18.7390$& &\\
   4 &$     1.690$ &$    -2.458$         &B &$  17.9223$  &I &$  16.9373$   &CFHT\\
\tableline\tableline
   1&$    94.010$&$  -378.603$&V&$  17.0280$&B&$  17.3258 $&CTIO\\
   1&$    94.010$&$  -378.603$&B&$  17.3298$&I&$  16.7495 $&CTIO\\
   2&$  -112.995$&$   307.413$&V&$  16.7213$&B&$  16.9203 $&CTIO\\
   2&$  -112.995$&$   307.413$&B&$  16.9223$&I&$  16.5535 $&CTIO\\
   3&$  -163.805$&$   -47.989$&V&$  17.1095$&B&$  17.5135 $&CTIO\\
   3&$  -163.805$&$   -47.989$&B&$  17.5144$&I&$  16.5490 $&CTIO\\
   4&$  -165.143$&$  -129.545$&V&$  17.4614$&B&$  17.7707 $&CTIO\\
   4&$  -165.143$&$  -129.545$&B&$  17.7746$&I&$  17.1466 $&CTIO\\

\tableline\tableline
\end{tabular}
\tablenotetext{a}{Coordinates are offsets from the center of the cluster}
\tablenotetext{b}{F5=F555W, F3=F336W, F2=F255W}
\tablecomments{The complete version of this table is in the
electronic edition of the Journal. The printed edition contains only a sample.}
\end{center}
\end{table}

\begin{table}
\begin{center}
\caption{New Horizontal Branch Star Table\label{hbtab}}
\begin{tabular}{crrcrcrc}
\tableline\tableline
Star & $\Delta \alpha$ (arcsec)\tablenotemark{a} & $\Delta \delta$ (arcsec)\tablenotemark{a} 
& Filter & Magnitude & Filter & Magnitude & Telescope \\
\tableline
1 &$-202.1681$  &$382.3869$  &$V$ &$15.8505$   &$B$ &$15.7814$  &CTIO\\
2 &$-307.5504$ &$-253.5796$  &$V$ &$16.2131$   &$B$ &$16.1191$  &CTIO\\
2 &$-307.5504$ &$-253.5796$  &$I$ &$16.4943$ & & &CTIO\\
3    &$4.7069$    &$1.1835$  &$V$ &$14.5378$   &$B$ &$14.2150$  &HST\\
3    &$4.7069$    &$1.1835$  &F555W &$14.4650$  & &         &HST\\
\tableline\tableline
\end{tabular}
\tablenotetext{a}{Coordinates are offsets from the center of the cluster}
\end{center}
\end{table}

\clearpage

\end{document}